\documentclass[12pt]{article}
\usepackage{times}
\usepackage{geometry}
\usepackage[utf8]{inputenc}
\usepackage{enumitem,amssymb}
\usepackage{ragged2e}
\usepackage{graphicx}
\usepackage{sidecap}
\usepackage[super,sort&compress]{natbib}

\newlist{thematic}{itemize}{8}
\setlist[thematic]{label=$\square$}
\usepackage{pifont}
\newcommand{\cmark}{\ding{51}}%
\newcommand{\done}{\rlap{$\square$}{\raisebox{2pt}{\large\hspace{1pt}\cmark}}%
\hspace{-2.5pt}}

\usepackage[superscript,biblabel]{cite}
\usepackage{float}

\usepackage{macros}
\usepackage{aas_macros}
\usepackage{wrapfig}

\usepackage{xcolor}

\begin{document}
{
\thispagestyle{empty} 
\raggedright
\huge
Astro2020 APC White Paper \linebreak
The NOAO Mid-Scale Observatories
\linebreak
\normalsize

\noindent \textbf{Thematic Areas:} \hspace*{60pt} \done Planetary Systems \hspace*{10pt} $\square$ Star and Planet Formation \hspace*{20pt}\linebreak
$\square$ Formation and Evolution of Compact Objects \hspace*{31pt} $\square$ Cosmology and Fundamental Physics \linebreak
  \done  Stars and Stellar Evolution \hspace*{1pt} \done Resolved Stellar Populations and their Environments \hspace*{40pt} \linebreak
  \done Galaxy Evolution   \hspace*{45pt} \done Multi-Messenger Astronomy and Astrophysics \hspace*{65pt} \linebreak
  
  \vspace{-0.05in}
\textbf{Authors:}
Lori Allen, Arjun Dey, Tim Abbott, Adam Bolton, Cesar Briceno, Jay Elias, Steve Heathcote, Jayadev Rajagopal, Abhijit Saha, Verne Smith
 \linebreak						
\textbf{Institution:} NOAO
 \linebreak
\textbf{Email:} lallen@noao,edu, adey@noao.edu, tabbot@noao.edu, abolton@noao.edu, cbriceno@noao.edu, jelias@noao.edu, sheathcote@noao.edu, jrajagopal@noao.edu, asaha@noao.edu, vsmith@noao.edu
 \linebreak


}
\smallskip
\noindent\textbf{Abstract:}
We describe present and future capabilities of the Mid-Scale Observatories (MSO) of the new national center merging NOAO, Gemini Observatory and LSST Operations. MSO is comprised of Cerro Tololo Interamerican Observatory (CTIO) and the Kitt Peak National Observatory (KPNO). Telescopes at both sites currently operate on a mix of public and private funding. Recent upgrades have equipped the MSO 4-m class telescopes to perform world-class surveys in diverse areas of astrophysics, from dark energy to exoplanets. 

\begin{figure}[b!]
\centering
\includegraphics[width=\textwidth]{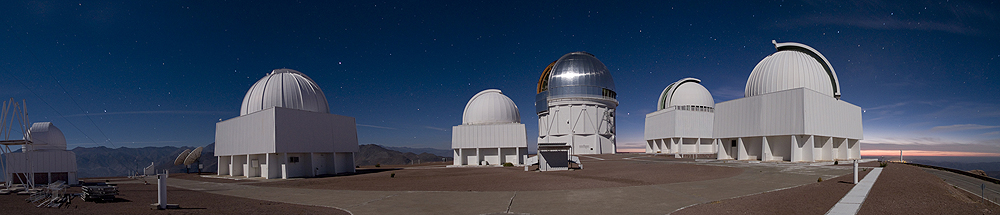}
\end{figure}

\begin{figure}[b!]
\centering
\includegraphics[width=\textwidth]{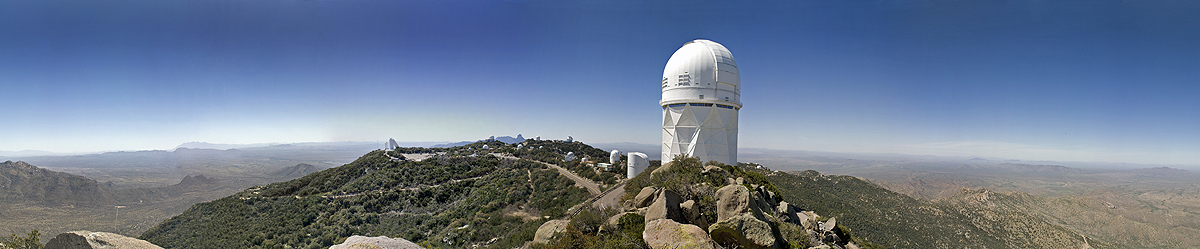}
\caption{The summits of Cerro Tololo (top) and Kitt Peak.}
\end{figure}

\pagebreak

\addtocounter{page}{-1}
\section{Key Science Goals: New Directions with Existing Optics}

The Kitt Peak National Observatory and the Cerro Tololo Inter-American Observatory will be consolidated under the Mid-Scale Observatories (MSO) division of the new organization created by the merger of NOAO, Gemini and LSST. The mission of MSO will continue to be to provide access to scientific research with world-class capabilities on 4-m class and smaller facilities for the US astronomical community. 

The facilities run by KPNO and CTIO remain critical to modern astronomical research. At the Blanco 4m telescope on Cerro Tololo, a partnership led by FermiLab resulted in the Dark Energy Camera, which has been transformational in the ability of the international astronomical community to undertake all-sky optical imaging surveys. The SOAR telescope is being transformed into a critical resource for the rapid response follow-up for astronomical transients. On Kitt Peak, the Mayall 4m telescope is currently being equipped with the Dark Energy Spectroscopic Instrument (DESI, funded by the Department of Energy), the world’s fastest machine for wide-field multi-object spectroscopic surveys, which will undertake the greatest cosmic cartography experiment of our time. And the WIYN telescope will soon host the NEID spectrograph (funded in partnership with NASA), which will be capable of making extremely precise radial velocity measurements of stars (rms$< 1~{\rm m\,s^{-1}}$) to identify and characterize exoplanets in our Galaxy. 

These telescopes and their instruments are at the cutting edge of US astronomy, and stand in stark contrast to the idea that “bigger is better”. The costs of new instrumentation have been $<$\$100M in all cases (for design, development, construction and installation), and operating costs remain $<$\$10M/yr for each facility. The manageable costs allow these facilities to be nimble and respond quickly to both changing science priorities and the exploration of new, perhaps more risky, technologies. 

Creating these capabilities, operating and maintaining them has necessitated changes in the operational model for the two observatories. The realities of the availability of funding for the operations of 4m-class telescope and the growing costs of cutting edge, capable astronomical instrumentation have required NOAO to partner with other funding agencies (public and private) to create these new opportunities. These partnerships have required dedicating the facilities to specific scientific projects for extended periods of time, in some cases with no other available access. This process has undoubtedly affected our traditional user base and left many US astronomers with limited access to these cutting edge facilities. 

With LSST and Euclid on the horizon, and with TESS, LIGO, ZTF and PanSTARRS currently producing exciting results by the day, the MSO telescopes will only increase in their relevance and impact. They will certainly be important as instruments to follow-up or complement observations made on other ground- or space-based telescopes. More importantly, operable at a fraction of the cost of larger facilities, they are critical as discovery tools, finding new sources for follow up by larger telescopes. The challenge is to secure funding for their operation through the next decade thereby not only guaranteeing access to the cutting edge instrumentation they already host, but also keeping that instrumentation current and relevant for the future.

The NSF has already invested significant capital in the construction, maintenance, and operation of these facilities over the last several decades.
Further investment at a modest level in these facilities in the coming decade can realize significant scientific benefits for the US astronomical community. Partnerships with other funding sources (e.g., DoE, NASA, private foundations) has and will continue to result in cutting edge instrumentation and world-beating capabilities.  Modest investments, perhaps for dedicated science campaigns or limited terms of operation, could ensure public access to these capabilities, resulting in significant benefit for small cost. 

\section{Technical Drivers: Multiple Modern Survey Platforms}

\subsection{Blanco+DECam}\label{blanco}

The Victor M. Blanco 4m telescope is more than 40 years old but because of its wide field optical, and rugged mechanical, design it has proven possible to modify, upgrade, and re-instrument it several times over its life and so maintain its position as a front-line research tool.  Most recently, in 2012 Blanco was substantially rebuilt to accommodate the Dark Energy Camera (DECam) - a high-performance, wide-field (3 square degree, 2.2-degree diameter), 520-megapixel CCD imager designed and built with funding from the Department of Energy (DOE). DECam has already completed data taking for two major multi-year surveys: The Dark Energy Survey (DES, 813 nights over 5.5 years) for which it was built, and the DECam Legacy Survey (DECaLS, 157 nights over 5 years). The balance of the telescope time has gone to a broad range of user-community science programs spanning the full gamut of science themes. A few of these are small PI programs, but several are major surveys in their own right. We anticipate that Blanco+DECam will continue as a forefront capability in heavy demand for the next decade and beyond and accordingly we are pursuing an active program of maintenance, renewal, and evolutionary upgrades to ensure its continued health. Over the next decade we expect that Blanco will:

\begin{itemize}
\item Continue to serve as an open-access imaging survey machine:  The completion of data collection for DES and DECaLS has opened up a substantial wedge of telescope time for new programs.  As the result of the strong response to a call-for-proposals issued in September 2019 about 60\% of the available time on Blanco over the next three years has been assigned to new medium- to large-sized survey programs. We expect to issue further survey calls to maintain the survey fraction at this level or higher as the current crop are completed. These programs include traditional surveys aimed at wide areal coverage, but also time-domain programs that repeatedly image smaller areas at cadences ranging from minutes to weeks. The remainder of the time is being allocated to smaller PI programs through the semi-annual calls for proposals. 

\item Become a factory for localization of the optical counterparts of gravitational wave (GW) sources and other rare transient events:  Although the upgraded LIGO/VIRGO which came back on line in April 2019 has somewhat smaller (but still  100 Deg$^2$) localization area, it also has greatly increased sensitivity allowing it to find significantly more distant GW sources. Thus the optical counterparts will typically be fainter than GW170817, itself likely atypical being both brighter and bluer than expected from theory.  Identification of such fainter counterparts against the background of more mundane transients such as SNe will require a systematic search of the entire localization area to faint limits. Since DECam can image a 100 Deg$^2$ search region to r,i  $\sim$23 in only 3 hours, it is a critical tool, unique in the southern hemisphere, for localization of  GW counterparts. DECam+Blanco is also being used in the search for counterparts of ICEcube neutrino events and for radio bursters. 

\item Transition to operations with an increasing emphasis on time-domain observing: MSO in collaboration with Las Cumbres Observatory (LCO), and Gemini is developing the Astronomical Event Observatory Network (AEON) to more effectively support time domain and target-of-opportunity science. AEON will support full dynamic scheduling with observations drawn from a pond based on priority.  PIs can add observations to the pond at any time, manually or based on output from an event broker. At Blanco the selected observations will be executed by trained service observers on site. This approach would support full time use of Blanco for transient follow-up.  But it can also mix in observations for traditional programs planned weeks or months in advance. Follow up of discoveries made with ZTF and other transient searches will allow end-to-end testing and optimization of our time domain infrastructure and policies in advance of the start of LSST observations while delivering high profile science. 

\item Play a key role in follow-up of LSST discoveries: Najita, Willman et al. (2016) call out wide field optical imaging on mid-scale telescopes as a high priority resource for diverse science cases, including studies of small solar system bodies, stellar rotation and activity, Milky Way science, and especially transients, and they note that DECam+Blanco would meet the needs for broad and medium-band imaging for many of these fields. Collocated with LSST, Blanco has immediate access to the same region of the sky and with DECam can match the depth of single visit LSST images in 180s in g-band and 240s in i- and z-bands. This will permit rapid follow-up of transient discoveries, allowing observation with the full filter set and/or filters not available in LSST or with different cadence (e.g. more frequent visits).
\end{itemize}
We are considering various renewal options that could help maintain the competitive edge of Blanco with costs that range from minor, to significant but still, very modest compared to a new facility.

\begin{itemize}
\item Additional filters for DECam cost ~US\$100k. Quite narrow band ($\delta\lambda/\lambda \sim$ 1-2\%) filters are now possible with good control of bandpass as a function of field radius. A filter centered on 964nm has been fabricated to identify Ly-$\alpha$ galaxies at the epoch of reionization (z=6.9; Zheng et al 2017 ApJ, 842, L22), and a red shifted H$\alpha$ filter has been successfully fabricated for another team.
\item Replacement of the focal plane arrays with new detectors e.g. Germanium CCDs, to extend the wavelength response. This would also require a new corrector. The cost would be comparable to but less than the build cost of the DECam instrument since much of the hardware could be reused.
\item As discussed in the ASTRO2020 white paper "Towards a Spectroscopic Survey Roadmap for the 2020s and Beyond" (Bolton et al.), DECam could be reconfigured as a wide field, massively multiplexed, optical spectrograph by installing a multi-fiber positioner behind the existing corrector.  This would have only about half the areal coverage of DESI, but could have almost as many fibers (4000 versus 5000) if tightly packed using the more compact Mohawk actuator design; this could feed clones of the DESI spectrographs.
\item With DECam mounted at prime focus, the Cassegrain focus on the Blanco is available to other instruments, and can accommodate, for example, the multi-object spectrograph COSMOS, or a large-field near-infrared imager like NEWFIRM. 
\end{itemize}

\subsection{Mayall/DESI}\label{mayall}

NOAO and the DESI Project have upgraded the Nicholas U. Mayall 4m telescope and are in the process of installing the DESI survey instrument at the telescope. Once commissioned later this calendar year, Mayall/DESI will represent the premier multiobject spectroscopic capability in the world.  With the ability to obtain spectra for 5000 targets simultaneously over a 3 degree diameter field of view, Mayall/DESI is the world’s fastest machine for very wide-field spectroscopic surveys. The initial DESI survey is funded through FY2026. For more information on Mayall/DESI capability, see the APC white paper "The Dark Energy Spectroscopic Instrument (DESI)" (Levi \& Allen, et al.).

The current model for support for this capability is that all the costs for operating the telescope and instrument are covered by the DESI Collaboration for the 5-year period of the survey. Data are pipeline processed and analyzed at the NERSC supercomputer facility at LBL and archived both at NERSC and at NOAO. The current plan is to make all the data public after a proprietary period.
Once the initial DESI survey is complete, the instrument may be available for future surveys. This second phase of DESI operations could be some combination of the following: 
\begin{itemize}
\item Dedicated long-term surveys (e.g., a survey of bright LAEs and Ly-$\alpha$ emitting LBGs to provide cosmological constraints at redshifts $\sim3<z<4.5$; survey of $\sim$10\% of Gaia stars for RV and spectral types, to constrain Galactic structure; high-density survey in the z$<0.1$ universe; AGN surveys to identify and monitor all central BHs in the LIGO-visible volume).
\item Public access surveys (e.g., 10-100 night surveys assigned through a peer-reviewed process)
\end{itemize}


Funding to support such future surveys would have to cover both the telescope and instrument operations as well as the pipeline processing of the data. Private collaborations could raise funds for dedicated surveys. Public access surveys would require an investment from the NSF to partially fund operations through the MSIP or similar program. Possible future directions for Mayall beyond the first 10 years of DESI could include:
\begin{itemize}
\item Upgrades to DESI (better/faster fiber positioners)

\item Different spectrographs - higher resolution, different wavelength coverage

\item New instrumentation (e.g., wide-field imaging camera behind DESI corrector, wide-field deployable IFUs, wide-field polarization).
\end{itemize}

\subsection{SOAR}\label{soar}
LSST will usher optical time-domain astronomy into a new and unprecedented dimension, bringing a scale of data management and mining not seen before. However, maximizing the scientific potential of LSST will require an efficient ecosystem of follow-up facilities (Najita, Willman et al. 2016). This has also been recognized by the National Research Council commissioned by the NSF. In their document "Optimizing the US Ground-based Optical and Infrared System", recommendations 2, 3 and 4 specifically instruct NSF to direct NOAO to implement and support ground-based capabilities that will optimize LSST follow-up science, and exploit the synergy between Gemini, SOAR, Blanco and LSST.

SOAR, being co-located with LSST, built to deliver excellent image quality over a reduced field of view, and equipped with a suite of instruments always available on its various ports, is particularly well suited to provide prompt follow-up observations for characterizing and further studying LSST targets, specially the brighter sources, over a wavelength range spanning from the UV to the near-IR.

\begin{figure}[h!]
    \centering
    \includegraphics[scale=0.5]{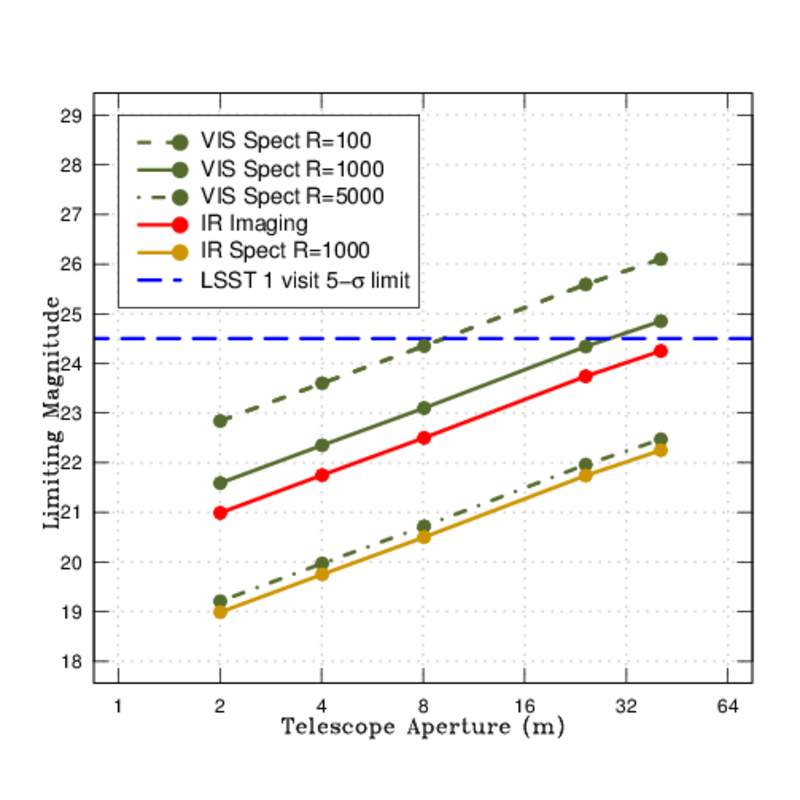}
    \caption{Sensitivity as a function of aperture for typical types of LSST follow-up observation. The LSST bright limit corresponds approximately to the lower edge of the figure. Data points in each curve represent 5-$\sigma$ limiting magnitudes for exposure times of 1h, except that at R=5000 the assumed S/N is 20. The horizontal blue dashed line is the approximate 5-$\sigma$ limiting magnitude across all bands for a single LSST visit, assuming 0.7$\arcsec$ seeing, which is equivalent to the limit reached with the DECam imager on the 4m Blanco telescope in 90s under similar conditions.}
    \label{fig:4m_sensitivities}
\end{figure}

Figure 2 puts the potential for a 4-m telescope like SOAR in context. It shows several plots of sensitivity vs. telescope aperture, for different instrumental configurations. Also shown on the plot is a typical sensitivity for LSST for a single visit. Also indicated is the LSST bright limit, above which photometry is not reliable.
The dashed box outlines the area where SOAR follow-up observations are potentially of value. For visible wavelength imaging, SOAR performance is similar to DECam on the Blanco telescope, though over a much smaller field of view. From the figure, it is evident that SOAR can perform useful follow-up observations over most or all of the LSST magnitude range with direct imaging and very-low-resolution spectroscopy. At higher spectral resolution or in the near-infrared, SOAR can work productively at the brighter end of the LSST magnitude range (or on redder objects in the near-infrared). Although such bright objects represent only a fraction of potential targets, they may well be the most interesting since they offer more opportunities for extended follow-up.

For SOAR to fully realize its potential for responsive follow-up to the LSST surveys, evolution of its operation is required. In particular, transition to queue operation for a substantial fraction of available observing time is considered critical, provided this can be done in a cost-effective manner. Furthermore, integration within a network continuing other telescopes (both of similar size and of different aperture) promises to use SOAR most efficiently. The prototype for this network is AEON (see section \ref{aeon} and Bryan Miller et al.'s APC white paper “Infrastructure and Strategies for Time Domain and MMA and Follow-Up”). 
These considerations are consistent with the recent recommendations of the AAAC sub-committee (2019).\footnote{https://www.nsf.gov/attachments/296041/public/aaac\_feb25\_klaus\_honschied.pdf}
That sub-committee also identified the need to provide additional instrumental capabilities to optimize rapid follow-up of interesting events, specifically upgrading or replacing the existing IR imager, and adding a low-resolution, very-wide-wavelength spectrograph (uv to NIR). Both these instruments, in particular the latter, would also be useful for classical “static” program which need to efficiently classify large samples of objects.
In order to pursue these objectives, SOAR must find resources outside of its regular operating budget; progress is likely to be resources-limited for this reason. The overall scope of these projects is, nonetheless, relatively modest, and would (for example) fit within the lower tier of the NSFI MSRI program.

\subsection{WIYN/NEID}

KPNO operates and maintains WIYN on behalf of the WIYN partnership, whose equity members are the University of Wisconsin, Indiana University, and NOAO, joined by operational partners University of Missouri-Columbia, Purdue University, and (as of 1 August 2019) Penn State University. 

Beginning in 2020, the NASA-NSF Exoplanet Observational Research (NN-EXPLORE) program will begin open-access community observing programs with the NN-EXPLORE Exoplanet Investigations with Doppler Spectroscopy (NEID) instrument on the WIYN 3.5-m telescope using NCOA/MSO time.  NEID (pronounced “noo-id”) is a new cutting-edge high-precision spectrograph at WIYN designed for radial velocity (RV) measurements of exoplanet host stars. It is designed with a goal of achieving 27 cm/s precision per data point, providing the US exoplanet community with high-precision RV measurements appropriate for studying Earth and super-Earth mass planets orbiting bright host stars over a wide range of spectral types.   More information on NEID can be found at:
 https://neid.psu.edu/what-is-neid/
  
Other investigations that hinge on extreme RV precision can also make good use of NEID.  In particular, NEID will fill in needs within the growing field of exoplanet and host-star characterization and will be a timely resource to support follow-up observations in the era of NASA’s TESS mission.  MSO will operate NEID in a queue-scheduled mode, and the NASA Exoplanet Science Institute (NExScI) will employ pipeline data reductions on all observations to provide PIs with high-level data products, including high-precision RVs.

The MSO share of observing time on the WIYN 3.5-m telescope is approximately 40\%, with this time available as open-access NN-EXPLORE time, thus WIYN plus NEID represents a significant asset to the exoplanet community into the decade of the 2020s.

\subsubsection{Other WIYN Instrumentation}

WIYN is equipped with a variety of instruments. These will continue to be available and supported throughout the NEID era, for both WIYN partners and NN-EXPLORE program observers. 
\begin{itemize}
\item One Degree Imager (ODI) — a wide-field optical imager with 40$\times$48 arc minute field of view with 0.11 arcsec pixels. The ODI focal plane is a 5$\times$6 mosaic of Orthogonal Transfer Arrays (OTAs).  The median delivered image quality (DIQ) in the r' band is 0.65 arcsec with a best DIQ of 0.35 arcsec in ideal conditions, fully exploiting the exceptional seeing at the WIYN site. Across its field of view, the ODI DIQ is competitive or better than every other ground-based imager in the world. Utilizing an atmospheric dispersion correction system and a full suite of 5 broadband and 4 narrow band filters, ODI delivers excellent image quality over the entire visible wavelength range. 

\item Hydra — a multi-object fiber spectrometer that is capable of observing up to 100 objects simultaneously across a 60-arcminute diameter field of view. Both blue and red optimized cables are available. Hydra feeds the Bench Spectrograph, which uses all-transmissive collimator and Virtual Phase Holographic (VPH) gratings to maximize throughput. Resolving powers in the range 1000–20,000 are possible at various central wavelengths.

\item SparsePak — a sparsely packed fiber optics bundle with nearly integral core that has a special fiber geometry designed to optimize performance for the specific scientific goal of studying the spatial distribution of the internal motions of gas and stars in nearby galaxies, but that is also useful in general for the study of galactic and extra-galactic nebulae. SparsePak feeds the Bench Spectrograph (see above).

\item WIYN High-Resolution Infrared Camera (WHIRC) — a near-infrared (0.9$-$2.5 $\mu$m) imager that installs on the WIYN Tip/Tilt module (WTTM) port. WHIRC has a 3.3 arcmin diameter field of view with 0.1 arcsec pixels and is capable of delivering near diffraction-limited images (~0.2 arcsec) in the K-band. It has a large selection of wide- and narrow-band filters.

\item NASA Exoplanet Star and Speckle Imager (NESSI) —  This instrument utilizes two electron-multiplying CCD cameras to capture speckle images in two colors simultaneously. Speckle observations provide resolutions at or near the diffraction limit of the telescope, allowing companion detection and characterization to delta magnitudes of approximately 5. NESSI has remote-controlled filter wheels in each beam, split by the dichroic at 686.4 nm. The EMCCDs can operate with high sensitivity and low noise even at very fast readout rates, providing high time resolution. 
\end{itemize} 

\subsection{Time Domain Follow up with AEON}\label{aeon}

The Astronomical Event Observatory Network (AEON) is the response to the need for a follow up ecosystem for the LSST era. Born out of a collaboration between NOAO, LCO, SOAR and Gemini, AEON has been built on the existing expertise of LCO in managing in an automated and with little human intervention, a distributed network of different telescopes with a variety of instrumentation. The 4m SOAR telescope, co-located with LSST on Cerro Pachon, with significant experience in remote observation and ongoing efforts for increased automation, was selected as the pathfinder facility for incorporating 4m-class facilities into the LCO network.  The 8m Gemini-South telescope is also actively taking steps to joining AEON once SOAR is fully operational within the network by 2020. AEON is aiming at providing users with a system that can respond quickly and efficiently to alerts, but at the same time also provide a wide spectrum of facilities for detailed studies of time-domain astronomy, and/or coordinated studies using several telescopes, including space-based facilities.  

AEON is part of an inter-connected system that includes brokers and Target Observation Managers (TOMs). Alert brokers like ANTARES filter data streams from ZTF or LSST, feeding targets to user-developed TOMs that have built into them the intelligence necessary to qualify targets that fit a particular science cases, such targets can then be submitted to specific facilities for follow up observations. A highly automated scheduling software that runs without human intervention is at the heart of AEON, it handles observation requests to each facility and receives feedback on whe.h r the observation was complettdeo  ronteDuring the 2019A semester SOAR has successfully run end-to-end tests of its custom-developed AEON software and interfaces, and is now starting shared-risk AEON observations during the 2019B semester. 

\subsection{Other MSO Telescopes}

The smaller ($D < 4$m) telescopes on both Cerro Tololo and Kitt Peak remain in use, operated largely by private$+$public consortia, under agreement with MSO. On both mountains, observatory staff maintain site-wide infrastructure "in between the domes", funded through cost recovery from the tenant observatories. The telescopes are engaged in a combination of research and training; some have been roboticized by their operators, some require on-site observers. 

These small telescopes are inexpensive to operate and enable long-term research projects, like time-domain follow-up and variability monitoring, to be carried out with little overhead. Some community open access time is currently available on some of the small (1-2m class) telescopes, and some existing consortia are welcoming new members. 

\section{Organization, Status and Schedule:The Long-Term Future of MSO}

The MSO assets (telescopes, operations and support expertise, and scientific staff) are invaluable to the progress and continued dominance of US ground-based OIR astronomy. Although much of the focus and expenditure of the funding agencies is dominated by the largest telescopes and new facilities (e.g., the US-ELT and LSST projects), the $\le$4-m class telescopes are critical for breakthrough science both by enabling discovery of new astrophysical horizons and for providing the much-needed complementary and supporting data for the larger-aperture telescopes. With projects like DECam, DESI and NEID on the Blanco, Mayall and WIYN telescopes, MSO will continue to provide the US community with powerful survey capabilities. 

While the lead time for constructing instruments in the DECam/DESI class is several years, these are now built, and new survey projects could be initiated without additional new hardware. The 4-m class survey machines (Blanco/DECam, Mayall/DESI, WIYN, SOAR) are cost-effective, reliable, and well understood. Without continued access to these facilities, the US community will not only miss out on great survey capabilities, but will also have to rely on follow-up of Multi-Messenger and Time-Domain discoveries from private or non-US facilities, a situation that could seriously impede scientific progress.  

Now is the time to start planning new projects for MSO telescopes in 2025 and  beyond. 

\subsection{Broader Impacts} 

It should be noted that the MSO telescopes perform an additional important role, which is training of graduate students and early-career scientists. A surprising number of graduate students and post-docs trained at MSO facilities have ended up in support or leadership roles at major observatories. Students who train on MSO telescopes and go on to faculty and research positions have a better understanding of what's required to plan and carry out observational experiments.   

In the public sphere, both Kitt Peak and Cerro Tololo are recognized brands that attract thousands of visitors to the observatories every year. It is hard to overstate the impact that a functioning professional research observatory has on interested visitors. Thus under MSO, both observatories will retain their names. 
Kitt Peak National Observatory is open during daylight hours to the public, free of charge, every day. School groups tour the observatory frequently, led by the excellent Docents of the KPNO Visitor Center. Public night-time observing programs are offered every clear night, for which an admission fee is charged, and during high season there is more demand than there is capacity. 

The high level of public interest in astronomy has resulted in a new outreach center, currently under construction on Kitt Peak, inside the former MacMath-Pierce Solar Telescope facility. The Windows on the Universe Center for Astronomy Outreach will provide a much-enhanced visitor experience, including interactive exhibits, a planetarium, and a Science-on-a-Sphere (SoS) theater. The entire center will be designed to bring visitors into close contact with NSF-funded, ground-based astronomy research from around the world. Scheduled to open in 2021, it is anticipated to attract several tens of thousands of patrons every year. In addition, through the SoS global science center network, results from U.S. observatories can be shared with other SoS theatres globally, vastly increasing the number of people reached. 

\pagebreak
\begin{figure}[ht!]
    \centering
    \includegraphics[width=0.9\textwidth]{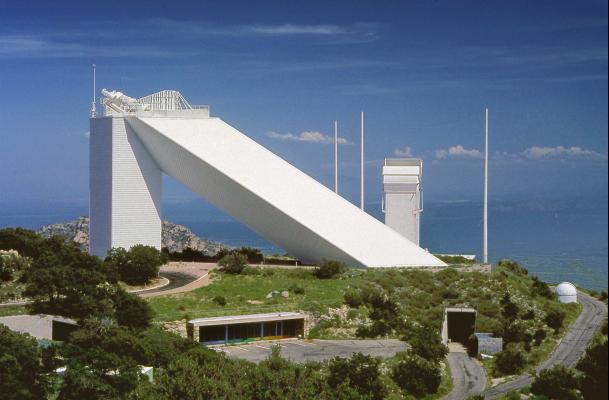}
   \caption{The McMath-Pierce solar telescope facility. Formerly the largest solar telescope in the world, it will soon be home to the Windows on the Universe Center for Astronomy Outreach. The (mostly underground) interior space is currently under renovation.}
\label{fig:mcmath}
\end{figure}
\section{Conclusion}

Recent investments in the Mid-Scale Observatory telescopes on Kitt Peak and Cerro Tololo have resulted in state-of-the-art, 4-m class survey machines that are unsurpassed in cost-effective performance. Thus we encourage the Astro2020 committee to recommend: 

1. Continued support and public access for the US community to existing capabilities on the MSO telescopes, particularly DESI, DECam and NEID,

2. Continued investment in new instrumentation at the \$10M-100M level on these facilities, and 

3. That the funding agencies encourage private-public and inter-agency partnerships to create new world-beating capabilities while ensuring some level of peer-reviewed public access to these capabilities for the US community.

\end{document}